\documentclass[prb,twocolumn,showpacs,superscriptaddress]{revtex4}

\usepackage{graphicx}

\newcommand{\ba}{\begin{eqnarray}}
\newcommand{\be}{\begin{equation}}
\newcommand{\ea}{\end{eqnarray}}
\newcommand{\ee}{\end{equation}}

\newcommand{\SLS}{S_{\mathrm{LS}}}
\newcommand{\SHS}{S_{\mathrm{HS}}}

\newcommand{\ignore}[1]{}

\begin{document}

\title{Ground-state phases in spin-crossover chains}

\author{Carsten Timm}
\email{timm@physik.fu-berlin.de}
\affiliation{Institut f\"ur Theoretische Physik, Freie Universit\"at Berlin,
Arnimallee 14, D-14195 Berlin, Germany}
\author{Ulrich Schollw\"ock}
\affiliation{Institut f\"ur Theoretische Physik C, RWTH Aachen, D-52056 Aachen,
Germany}

\date{October 25, 2004}

\begin{abstract}
Spin-crossover molecules having a low-spin ground state
and a low lying excited high-spin state are promising components for
molecular electronics. We theoretically examine one-dimensional
spin-crossover chain molecules of the type of $\mathrm{Fe}^{2+}$ triazole
complexes. The existence of the additional
low-spin/high-spin degree of freedom leads to
rich behavior already in the ground state. We obtain the complete
ground-state phase diagram, taking into account an elastic nearest-neighbor
interaction, a ferromagnetic or antiferromagnetic exchange interaction
between the magnetic ions, and an external magnetic field.
Ground-state energies are calculated with high numerical precision
using the density-matrix renormalization group (DMRG). Besides pure
low-spin, high-spin, and alternating low-spin/high-spin phases we
obtain a number of periodic ground states with longer periods, which we
discuss in detail. For example, for antiferromagnetic coupling there exists
a dimer phase with a magnetic unit cell containing
two high-spin ions forming a spin singlet and a single low-spin ion, which
is stabilized by the energy gain for singlet formation.
\end{abstract}


\maketitle

\section{Introduction}
\label{sec.intro}

One of the most active fields of materials science to emerge in recent years is
\emph{molecular electronics},\cite{JGA00,NiR03} which proposes to use individual
molecules as electronic components. A related idea is to use a single quantum
spin, say of an ion, to store information. Kahn and
coworkers\cite{Kahn,BoK95,KaM98} have emphasized that \emph{spin-crossover
compounds} \cite{CaS31,Bonding1,Bonding2,GGG00,BlP04} (SCC's) are particularly
promising for molecular memory devices. 
These compounds consist of complexes involving transition-metal ions and organic
ligands.\cite{Bonding1,Bonding2,GGG00,BlP04} The magnetic ions can be either in
a low-spin (LS) or high-spin (HS) state, i.e., for the spin operator
$\mathbf{S}_i$ at site $i$ the eigenvalues of $\mathbf{S}_i\cdot\mathbf{S}_i$
are $\SLS(\SLS+1)$ and $\SHS(\SHS+1)$ in the LS and HS state, respectively. The
energy difference between HS and LS states is due to the competition between the
crystal field splitting, which prefers doubly occupied d-orbitals and hence LS,
and Hund's first rule, which favors the HS
state.\cite{Bonding1,Bonding2,GGG00,BlP04} In SCC's the LS state is the ground
state and the HS state is at a moderate (thermal) excitation energy. SCC's show
a characteristic crossover from the LS ground state to dominantly HS behavior at
higher temperatures\cite{CaS31} due to the higher degeneracy of the HS state.
This crossover is typically sharper than expected for noninteracting magnetic
ions and is even replaced by a first-order transition in several
compounds.\cite{Bonding1,Bonding2,GGG00,BlP04} Spin-crossover phenomena are also
observed in organic radicals\cite{radical} and certain inorganic
transition-metal compounds.\cite{Co}

Of the large number of known SCC's some naturally form one-dimensional chains,
for example $\mathrm{Fe}^{2+}$ with 4-R-1,2,4-triazole
ligands.\cite{KaM98,YMK98} Three ligands form bridges between two adjacent iron
ions. Other SCC's consist of two-dimensional layers, for example
TlSr$_2$CoO$_5$.\cite{Co,KhL04}

Besides possible applications,\cite{Kahn,BoK95,KaM98} SCC's are also interesting
from a statistical-physics point of view. Compared to conventional
local-moment systems they introduce an additional Ising degree of freedom
$\sigma_i$, which destinguishes between the LS ($\sigma_i=+1$) and HS
($\sigma_i=-1$) states. In the case of a \emph{diamagnetic} LS state ($\SLS=0$)
the low spins are essentially switched off. These SCC's are thus related to
site-diluted spin models,\cite{VMG02,dilute,YRH04} but in our case the presence
or absence of a spin is a \emph{dynamical} variable and not a type of
quenched disorder. Also related are recent studies of magnetic models with
mobile vacancies\cite{HLS04} and of insulating phases of atoms with spin
in optical lattices.\cite{ILD03} We show below that there is also a
close relation to finite antiferromagnetic spin chains.

Antiferromagnetic spin chains have attracted a lot of interest since Haldane's
(in the meantime firmly established) conjecture of a fundamental difference
between (isotropic) half-integer and integer quantum spin chains.\cite{Hald82}
Among other things, the latter always show an excitation gap, while the former
are critical. The valence-bond-solid model (AKLT model\cite{Affl87}), in which
each spin of length $S$ is replaced by $2S$ fully symmetrized spin-1/2 objects
that are then linked by singlet bonds between sites, was found to explain all
main features of integer quantum spin chains. One peculiarity of the AKLT model
is that at each end of \emph{open} spin chains $S$ of the spin-1/2 objects find
no singlet partner and form a free spin $S/2$. For integer spins this leads in
the AKLT model for \emph{even} chain lengths to a $[2(S/2)+1]^2 = (S+1)^2$-fold
degenerate ground state instead of the non-degenerate ground state found for
periodic boundary conditions. This observation carries over to antiferromagnetic
Heisenberg chains. There, one finds a group of $(S+1)^2$ low-lying states that
become degenerate exponentially fast for long open chains. The lowest-lying of
these states has total spin 0; above this state there follows a spin-1 triplet,
etc. The maximum total spin in this group of states is given by $S$ and is
concentrated at the edges. The lowest-lying excitation above them is the first
true bulk excitation and corresponds to the lowest-lying state with $M=S+1$.
This phenomenon has been observed experimentally,\cite{Hagi90} and generates a
wealth of low-lying excitations if there are segments of spin chains of various
lengths like in SCC's with $\SLS=0$. For \emph{odd} chain lengths, the situation
is different as the lowest-lying
states of the magnetization sectors $M \leq S$ are strictly degenerate both in
the AKLT and Heisenberg models. However, the ground state of the magnetization
$S+1$ sector again contains the lowest-lying bulk excitation. This observation
points to a special role of magnetization $M=S$, as will be seen throughout this
paper.

In the present paper we focus on the ground-state properties of one-dimensional
SCC's, for which we obtain essentially exact results. In particular, we treat
the case of a diamagnetic LS state appropriate for $\mathrm{Fe}^{2+}$ ions.
We mostly consider \emph{antiferromagnetic} coupling between the spins,
which is probably the more common situation.

\section{Theory}

We start from the Hamiltonian\cite{GKA00}
\be
H_0 = -V \sum_{\langle ij\rangle} \sigma_i \sigma_j - B_0 \sum_i \sigma_i
  - h \sum_i S_i^z .
\label{eq3.H02}
\ee
The sum over $\langle ij\rangle$ counts all nearest-neighbor bonds
once and the eigenvalues of $S_i^z$ are $m_i=-S_i,-S_i+1,\ldots,S_i$, where
$S_i=\SLS$ ($\SHS$) for $\sigma_i=1$ ($-1$). $V$ describes an interaction that
for $V>0$ ($V<0$) favors homogeneous (alternating) arrangements of LS and
HS. At least in a subset of known SCC's this interaction is
of \emph{elastic} origin\cite{elastic} and can be of either sign.\cite{KhL04}
We approximate this interaction by a nearest-neighbor term.
$2B_0>0$ describes the energy difference between HS
and LS and $h$ is the physical magnetic field with $g$ factor and Bohr magneton
absorbed. $H_0$ is diagonal in
the basis of eigenstates of all
$\mathbf{S}_i\cdot\mathbf{S}_i$ and $S_i^z$. In this basis\cite{rem.sigmaop}
\be
H_0 = -V \sum_{\langle ij\rangle} \sigma_i \sigma_j - B_0 \sum_i \sigma_i
  - h \sum_i m_i .
\label{eq3a.H02}
\ee
For $h=0$ we
reobtain the Ising-type model introduced by Wajn\-flasz and Pick\cite{WaP70} for
magnetic molecular compounds and by Doniach\cite{Don78} for lipidic chains.
Here, each site can be in two states characterized by $\sigma_i$ like in the
Ising model, but the states are degenerate with degeneracies $2\SLS+1$ and
$2\SHS+1$.\cite{rem.degen} The model can be rigorously mapped
onto an Ising model in a temperature-dependent effective field
$B$\cite{Don78,NMI96,GKA00} and has been treated in the mean-field
approximation
and with Monte Carlo simulations.\cite{BoK95} A
related model with next-nearest-neighbor elastic
interactions has recently been studied by Monte Carlo simulations
and a number of stripe phases have been found.\cite{KhL04}
The one-dimensional model
suitable for triazole compounds has not been treated before.

We are interested in a model with an additional exchange interaction $J\neq 0$
between the magnetic ions. The Hamiltonian (\ref{eq3.H02}) is generalized to
\be
H = H_0
  - J \sum_{\langle ij\rangle} \mathbf{S}_i \cdot \mathbf{S}_j ,
\label{eq4.H02}
\ee
where $J>0$ ($J<0$) corresponds to a ferromagnetic (antiferromagnetic)
coupling between the spins.
$J$ has not been measured in SCC's, but it has been
determined in similar metal-or\-ga\-nic complexes. $|J|/k_B$ is typically of
the order of $10$ to $20\,\mathrm{K}$ and is
antiferromagnetic,\cite{WHM97,BSC04,LBP04} as expected for a kinetic
superexchange interaction.\cite{Anderson}
In compounds based on Prussian blue, larger exchange interactions have
been observed, leading to ferrimagnetic order
at room temperature.\cite{FMO95}

The full Hamiltonian $H$ commutes with the operators
$\mathbf{S}_i\cdot\mathbf{S}_i$. Thus the total spin \emph{at each site} is
a constant of motion. On the other hand, $H$ does no longer commute
with $S_i^z$. In the case of $\mathrm{Fe}^{2+}$, which is the most common
magnetic ion in SCC's, we have $\SHS=2$ and $\SLS=0$ and a significant
simplification ensues, since
any low spin partitions the chain into finite segments that do not interact
\emph{magnetically}. Thus there is a close relation to the physics of finite
spin chains. In more than one dimension $\SLS=0$ leads to a less trivial
percolation problem---for long-range order to be present it is necessary
for the high spins to percolate.

In the following we restrict ourselves to $\SLS=0$.
The Hamiltonian can be written as
\ba
H & = & -V \sum_i (\sigma_i\sigma_{i+1}-1)
  - B_0 \sum_i (\sigma_i - 1) \nonumber \\
& & {}- J \sum_i \mathbf{S}_i\cdot \mathbf{S}_{i+1}
  - h \sum_i S^z_i ,
\label{eq5.H02}
\ea
where we have added a constant so that
the energy of the pure LS state vanishes. The Ising
operators $\sigma_i$ all commute with the Hamiltonian $H$. Their eigenvalues
are thus good quantum numbers and the Hilbert space is a direct product
of subspaces for given
$\{\ldots,\sigma_i,\sigma_{i+1},\ldots\}$.\cite{rem.sigmaop}

In each sector $\{\ldots,\sigma_i,\sigma_{i+1},\ldots\}$ the system
consists of chains of high spins separated by chains of low spins. The pure
HS and LS states are obtained as the obvious limits. Since
the LS chains do not contribute to the energy,
the total energy in a sector can be written as a sum over the energies
of HS chains of various lengths, including a contribution from their ends.
These HS chains do not interact magnetically since $\SLS=0$.

We are interested in the ground state and thus consider the \emph{lowest}
energy in each sector. The lowest energy in a sector can be written as
the sum over the ground-state energies of non-interacting finite HS chains.
Since $H$ commutes with the total spin of each
HS chain separately, the $z$-components
$M$ of the total spins of the finite chains are good quantum
numbers. Let us denote the lowest energy of a HS chain of
length $n$ with magnetic quantum number $M$ by $e_n^0(M)$, where
$|M|\le n\SHS$. We write
\be
e_n^0(M) = 4V + 2n B_0 - hM + \Delta e_n^0(M) ,
\label{eq5.e2}
\ee
where the first term comes from the extra energy of the change from HS to LS
at the ends. The final term is the lowest eigenenergy
of the finite HS Heisenberg chain with open boundary conditions and the
Hamiltonian
$H_n = -J \sum_{i=1}^{n-1} \mathbf{S}_i \cdot \mathbf{S}_{i+1}$.

\section{Results and Discussion}

\subsection{Ferromagnetic Coupling}

For ferromagnetic coupling, $J>0$, and magnetic field
$h>0$ both the exchange interaction and the Zeeman term favor ferromagnetic
alignment. The lowest-energy state thus has the
maximum magnetic quantum number $M=n\SHS$ for each chain.
For $J\ge 0$ we have $\Delta e_n^0(n\SHS) = -(n-1)\, J\SHS^2$ and thus
\be
e_n^0(n\SHS) = 4V + 2n B_0 - n h \SHS - (n-1)\, J \SHS^2 .
\ee
Now let us consider a sector $\{\ldots,\sigma_i,\sigma_{i+1},\ldots\}$ for
which the state consists of \emph{volume fractions} $p_n$ of HS chains of
length $n$. The HS chains have to be separated by at least
one low spin. When counting this low spin with each HS chain,
the volume fractions become $(n+1)/n\, p_n$. They must
satisfy the constraint
\be
\sum_{n=1}^\infty \frac{n+1}{n}\, p_n \le 1 .
\label{eq5.cstr2}
\ee
The energy per site is
\ba
\epsilon_0 & = & \sum_n p_n\, \frac{e_n^0(n\SHS)}{n} \nonumber \\
& = &
  \sum_n p_n\, \left(\frac{4V+J\SHS^2}{n} + 2B_0 - h\SHS - J \SHS^2\right)
  .\quad
\ea
We have to solve the linear optimization problem of minimizing
$\epsilon_0$ under the constraint (\ref{eq5.cstr2}).
In the space of vectors $(p_1,p_2,\ldots)$ the
region allowed by Eq.~(\ref{eq5.cstr2}) is a hyperpyramid with apex at
zero and the other corners
at points with $p_m=m/(m+1)$ for one $m$ and $p_n=0$ for $n\neq m$.
These are the points for which only chains of one single
length are present and have the maximum volume fraction.
This means that the finite
HS chains are separated by \emph{single} low spins.
Since the allowed region is convex, the only possible
solutions are its corners, except for special choices of parameters. Thus
either $p_n=0$ for all $n$ (LS state) or
$p_m=m/(m+1)$ for one $m$ and all other $p_n=0$.
For the LS state we have $\epsilon_0=0$,
whereas for the state with nonzero $p_n$,
\be
\epsilon_0
  = \frac{4V\!-2B_0+2J\SHS^2+h\SHS}{n+1} + 2B_0 - h\SHS - J \SHS^2 .
\label{eq5.eps04}
\ee
Examination shows that there are only three possible phases:
(i) If $4V-2B_0+2J\SHS^2+h\SHS>0$ and 
$2B_0-h\SHS-J\SHS^2>0$ or $4V-2B_0+2J\SHS^2+h\SHS<0$ and
$2V + B_0 - h\SHS/2>0$ the ground state is the LS state.
(ii) If $4V-2B_0+2J\SHS^2+h\SHS>0$ and $2B_0-h\SHS-J\SHS^2<0$
the ground state has $p_n>0$ and all other $p_m=0$,
for $n\to \infty$, which corresponds to the HS
state, and the energy is $\epsilon_0=2B_0 - h\SHS - J \SHS^2$.
Note that the HS state appears for any values of $V$ and $h$ for
sufficiently large exchange interaction $J$. This is reminicent of the
exchange-induced Van-Vleck ferromagnetism in rare-earth
compounds.\cite{FuP72}
(iii) If $4V-2B_0+2J\SHS^2+h\SHS<0$ and $2V + B_0 - h\SHS/2<0$ the ground state
has $p_1=1/2$ and all other $p_m=0$. This corresponds to an \emph{alternating}
state of low and high spins. The energy is $\epsilon_0 = 2V + B_0 -
{h\SHS}/{2}$.
By using the LS/HS splitting $B_0$ as our unit of energy, we obtain the
phase diagram in $V/B_0$, $h/B_0$, and $J/B_0$
shown as the $J\ge 0$ part of Fig.~\ref{fig.PD3D}, below.

\subsection{Antiferromagnetic Coupling}

\begin{figure}[tb]
\centerline{\includegraphics[width=3.35in,clip]{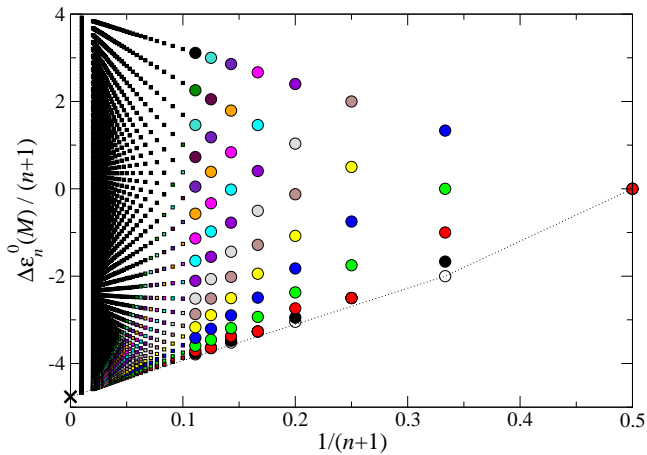}}
\caption{(Color online). The lowest energies $\Delta e_n^0(M)/|J|$ of
antiferromagnetic Heisenberg chains of length $n$ for spin $\SHS=2$ in
sectors with open boundary conditions for fixed total $S^z$ quantum number
$M$. The energies are normalized by a factor $1/(n+1)$ and shown as a
function of $1/(n+1)$, where $n+1$ is the period of the LS/HS pattern.
Circles: Results from
exact Lanczos diagonalization. Squares: Results from
DMRG. Equal colors correspond to equal $M$. For odd $n$ the
energies for $M=0$, $1$, and $2$ are degenerate,
as noted in Sec.~\protect\ref{sec.intro}. The
cross at $1/(n+1)=0$ denotes the extrapolated energy density $-4.761248(1)$
of an infinite chain.\cite{SGJ96}}
\label{fig.cheng}
\end{figure}

In the case of antiferromagnetic coupling, $J<0$,
there is a competition between the exchange and Zeeman terms in
Eq.~(\ref{eq5.H02}). Thus in principle finite HS chains of length
$n$ can occur with any magnetic quantum number $M$. The energy
$e_n^0(M)$ of such a chain is given by Eq.~(\ref{eq5.e2}). We introduce
volume fractions $p_{n,M}$ of HS chains of length $n$ with magnetic quantum
number $M$. They must satisfy the constraint
\be
\sum_{n=1}^\infty \sum_{M=-n\SHS}^{n\SHS} \!\! \frac{n+1}{n}\, p_{n,M}
  \le 1 .
\label{eq5.const4}
\ee
The energy per site is
\be
\epsilon_0 = \sum_{n,M} p_{n,M} \left( \frac{4V-hM+\Delta e_n^0(M)}{n}
  + 2B_0 \right) .
\ee
{}The ground state for certain parameter values is determined by the minimum of
$\epsilon_0$ under the constraint (\ref{eq5.const4}). This
is again a linear optimization
problem. Except for accidental degeneracies, the minima occur at the
corners of the allowed parameter region. Thus
either $p_{n,M}=0$ for all $(n,M)$ (LS state) or
$p_{n,M} = n/(n+1)$ for one $(n,M)$ and $p_{n,M}=0$ for all others.
In the latter case the energy per site is
\be
\epsilon_0 = \frac{4V-2B_0-hM+\Delta e_n^0(M)}{n+1} + 2B_0 .
\label{eq5.eps06}
\ee
$\Delta e_n^0(M)/|J|$ is the lowest energy of the finite antiferromagnetic
Heisenberg chain with open boundary conditions and the Hamiltonian $H_n' =
\sum_{i=1}^{n-1} \mathbf{S}_i \cdot \mathbf{S}_{i+1}$ in the sector with total
$S^z$ quantum number $M$. It is not possible to find these energies in
analytical form. For sufficiently small $n$, the Hamiltonian $H_n'$ can be
diagonalized numerically.
We have calculated the energies up to $n=8$ for all $M$ using
the Lanczos algorithm. The results for $\Delta e_n^0(M)/|J|/(n+1)$ are shown in
Fig.~\ref{fig.cheng} as colored circles, where identical colors denote the
same values of $M$.

The energies for longer chains can be calculated with excellent precision
with a finite-chain DMRG al\-go\-rithm;\cite{Whit92,Whit93} for a detailed
explanation of the algorithm and its applications see Refs.\ 
\onlinecite{Pesc99,Scho04}. To obtain typically seven-digit precision for
the ground state energy per site, we have kept up to $M=300$ states in
the reduced DMRG Hilbert spaces and carried out three
finite-system sweeps which was enough to ensure convergence. Note that
DMRG prefers open to periodic boundary conditions. In
standard DMRG applications to integer-spin chains, end spins of length $S/2$
(a spin 1 at each end for our case) are attached to eliminate the
peculiar boundary degrees of freedom and access bulk physics 
directly.\cite{Whit92,Scho95}
In the present calculation, these boundary degrees of freedom are physical
and hence no end spins are attached.
The energies for lengths $n=9$ through $n=49$ as well as $n=99$ are shown in
Fig.~\ref{fig.cheng} by squares.\cite{rem.largeM}

\begin{figure}[tbh]
\centerline{\includegraphics[width=3.40in]{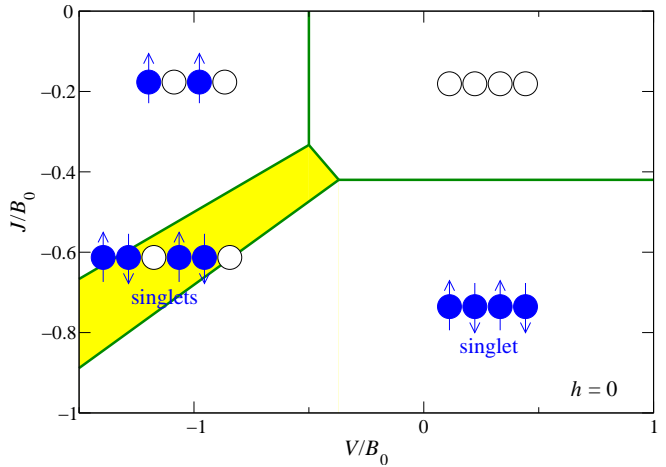}}
\caption{(Color online). Ground-state phase diagram of the 1D
spin-crossover model with $\SLS=0$ and $\SHS=2$ for vanishing magnetic
field, $h=0$, and antiferromagnetic exchange interaction, $J<0$. The dimer
($n=2$) phase case is highlighted. The heavy solid lines denote
discontinuous transitions. The various spin structures are indicated by
cartoons (solid symbols: HS state, open symbols: LS state); these should not
be overinterpreted---there is no magnetic long-range order.}
\label{fig.PDAFM1}
\end{figure}

We first consider the case of \emph{vanishing magnetic field}, $h=0$.
Figure~\ref{fig.cheng} shows that the state with $M=0$ has the lowest
energy for any $n$. Then $\epsilon_0 = ({4V-2B_0+\Delta e_n^0(0)})/({n+1})
+ 2B_0$ is a sum of $\Delta e_n^0(0)/(n+1)$ and a \emph{linear} function in
$1/(n+1)$. The minimum of $\epsilon_0$ can only occur for $n=1$, $n=2$, or
$n\to\infty$, since all other points lie above the dotted straight lines
connecting the corresponding points in Fig.~\ref{fig.cheng} (not obvious on
this scale). The relevant energies per site are thus determined by $\Delta
e_1^0/2=0$, $\Delta e_2^0/3=2J$, and $\lim_{n\to\infty}\Delta
e_n^0/(n+1)=4.761248\,J$ and have to be compared to the LS energy
$\epsilon_0=0$. The resulting phase diagram is shown in
Fig.~\ref{fig.PDAFM1}. Note the appearance of a \emph{dimer} ($n=2$) phase.
In this phase the energy increase due to the HS-HS pairs ($V<0$
favors HS-LS neighbors) is overcompensated by the large
negative singlet formation energy of Heisenberg spin pairs.

For \emph{general magnetic field} $h$ we have to take all possible magnetic
quantum numbers $M$ of the chains into account. This is obviously impossible for
the pure HS phase. Instead, we have performed DMRG calculations for chain length
$n=99$ for all possible magnetizations $M=0,\ldots,198$ and use them as a
caricature of the HS state. The resulting errors are discussed below.

\begin{figure}[tbh]
\centerline{\includegraphics[width=3.40in]{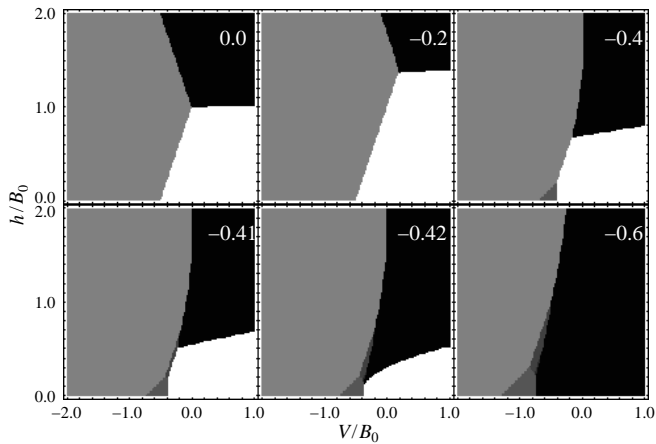}}
\caption{Zero-temperature phase diagrams for the same model as in
Fig.~\protect\ref{fig.PDAFM1}, but in a magnetic field, for
antiferromagnetic exchange interactions $J/B_0 = 0.0$, $-0.2$, $-0.4$,
$-0.41$, $-0.42$, $-0.6$. The white area corresponds to the LS phase, the
black to the HS phase, approximated by a phase with $n=99$, and the gray
areas correspond to $n=1, 2, 3, 5$ (from light to dark). All
transition are discontinuous, the purely magnetic continuous transition
discussed below is not shown.}
\label{fig.PDAFM2}
\end{figure}

For each set of parameters $(V/B_0,J/B_0,h/B_0)$ we
calculate the minimum energy densities for all states with $n\le 49$ as
well as $n=99$ from Eq.~(\ref{eq5.eps06}). The energy density of the LS
state is zero. Then the ground state is obtained by finding the minimum
energy. Figure \ref{fig.PDAFM2} shows a series of phase diagrams
for fixed exchange interaction $J$. Note that the lower edges of each
diagram, i.e., $h=0$, are consistent with Fig.~\ref{fig.PDAFM1}. We observe
that the dimer ($n=2$) phase present at $h=0$ is suppressed by the field,
as is expected since this phase is stabilized by the singlet formation
energy.

\begin{figure}[tbh]
\centerline{\includegraphics[width=1.60in]{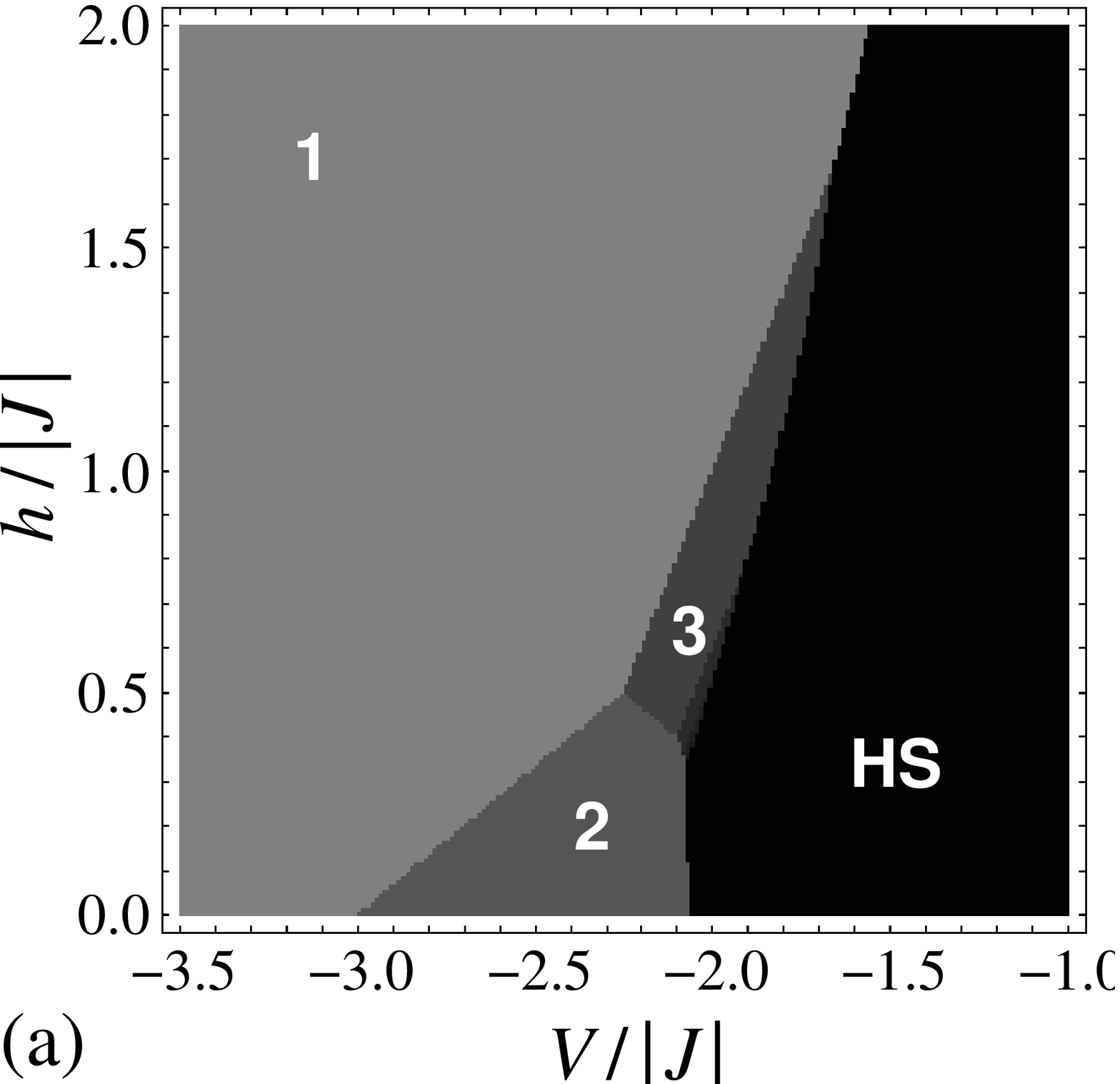}\hspace*{0.10in}\includegraphics[width=1.65in]{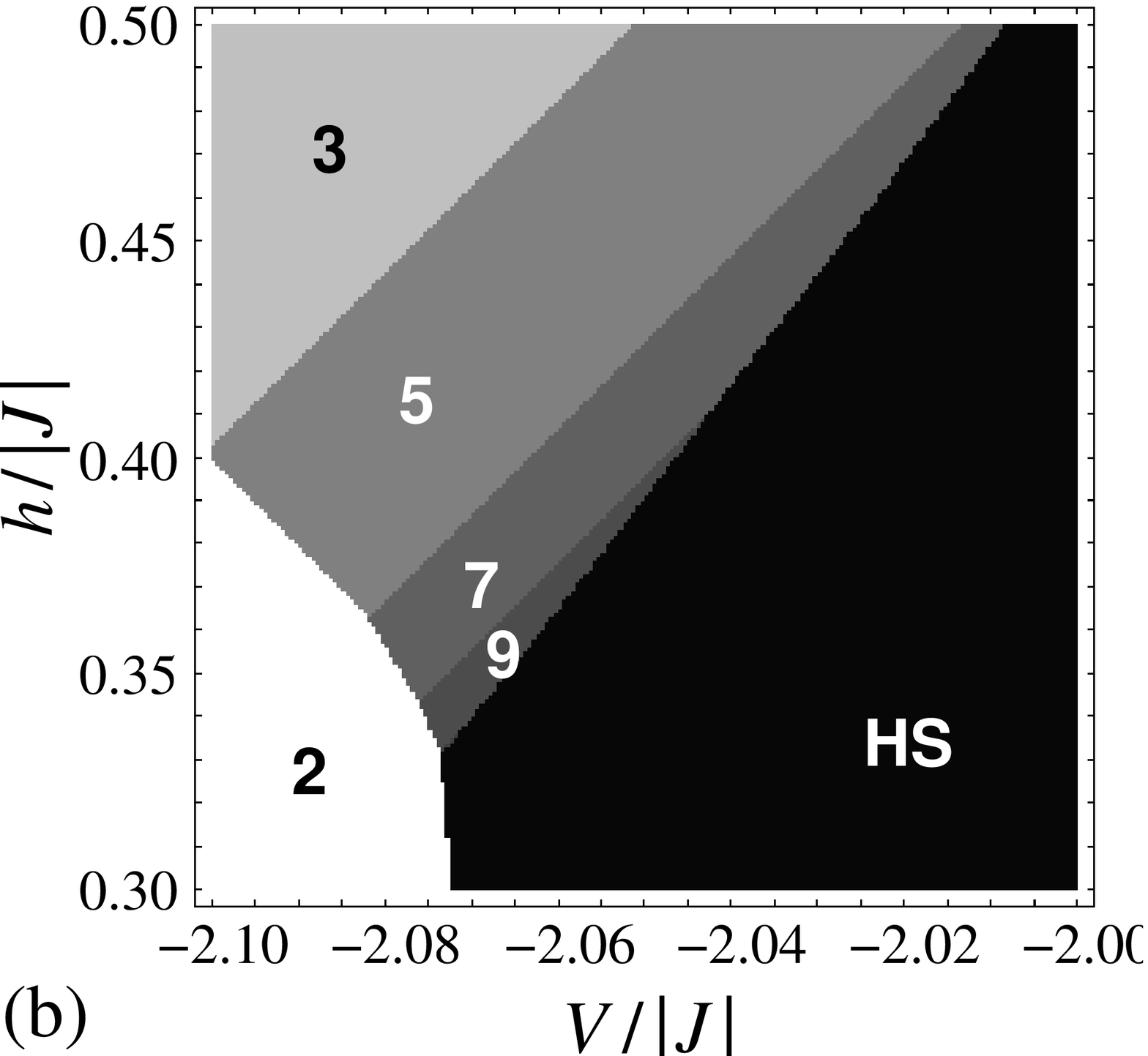}}
\caption{(a) Zero-temperature phase diagrams as in
Fig.~\protect\ref{fig.PDAFM2} in the limit $B_0\to 0$ (or $V, J,
h\to\infty$ with their ratios fixed). The gray scale is the same as in
Fig.~\protect\ref{fig.PDAFM2}. (b) Enlargement of the left figure on a
\emph{different gray scale}. The values of $n$ in the various ground states
are indicated.}
\label{fig.PDAFM3}
\end{figure}

For $J\lesssim -0.6$ the phase diagrams remain qualitatively the same. The
features are shifted to lower $V$ and expanded linearly in both the $V$ and
$h$ directions. Letting $V$, $J$, and $h$ go to infinity while keeping
their ratios fixed corresponds to the limit $B_0\to 0$, i.e., vanishing
energy difference between LS and HS states. In this limit we choose $|J|$
as our unit of energy, leaving two dimensionless parameters $V/|J|$ and
$h/|J|$. The resulting phase diagram is shown in Fig.~\ref{fig.PDAFM3}.

Interesting behavior is seen in the triangular region surrounded by phases with
$n=1$, $n=2$, and the HS phase, as shown in Fig.~\ref{fig.PDAFM3}(b). Here,
phases with HS chain lengths $n=3$, $5$, $7$, and $9$ are found. We do not
observe any further phases. To understand why only \emph{odd} $n$ appear, we
refer to the energy densities in Fig.~\ref{fig.cheng}: The energy of the singlet
($M=0$) state is \emph{lower} for even $n$ than expected from a linear fit, at
least for the small $n$ relevant here, while for odd $n$ it is higher. Thus at
zero magnetic field, even-$n$ states are preferred. On the other hand, the
energy of states with $M\ge 2$ is \emph{higher} for even $n$ than a linear fit,
while for odd $n$ it is lower. Thus in a sufficiently large magnetic field
odd-$n$ states are preferred. The fact that the series of odd $n$ is cut off at
$n=9$ is a result of the detailed numerical values of energies in
Fig.~\ref{fig.cheng}---for larger $n$, the HS state happens to have the lower
energy. While we expect the appearance of only odd HS chain lengths to be a
robust feature of spin-crossover chains with strong antiferromagnetic
interaction, the restriction to $n\le 9$ should thus be model dependent. For
example, inclusion of a next-nearest-neighbor elastic interaction or a different
integer value of $\SHS$ may change this result.

\begin{figure}[tbh]
\centerline{\includegraphics[width=3.40in]{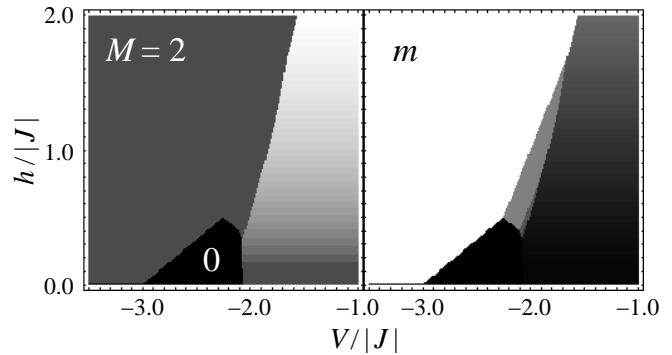}}
\caption{Left: Density plot of the magnetic quantum
number $M$ of the finite chains in the same parameter
region as in Fig.~\protect\ref{fig.PDAFM3}(a).
Black corresponds to $M=0$, white to maximum $M$. The phases with odd $n$
all have $M=2$. Right: Magnetization $m=M/(n+1)$ derived from the data in
the left plot. Black (white) corresponds to $m=0$ ($m=1$). The magnetization
of the fully polarized HS state, which does not appear in the plot,
would be $m=\SHS=2$.}
\label{fig.PDAFM4}
\end{figure}

\subsection{Magnetic Properties}

\begin{figure}[tbh]
\centerline{\includegraphics[width=3.35in,clip]{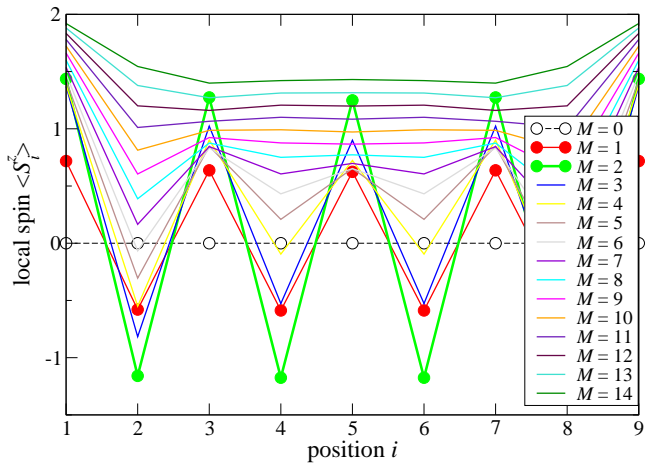}}
\caption{(Color online). Local expectation values
$\langle S_i^z\rangle$ from DMRG for each site of a
HS chain of length $n=9$ for various total magnetic quantum numbers of the
chain, $M$.}
\label{fig.local9}
\end{figure}

We now discuss the magnetic properties in more detail. In the ferromagnetic
case all high spins are fully aligned with a nonzero magnetic field $h$.
For the antiferromagnetic case Fig.~\ref{fig.PDAFM4} shows the magnetic
quantum number $M$ of the finite HS chains and the magnetization $m$ in the
ground states in the limit of large $V$, $J$, $h$. The magnetization is
defined as the magnetic quantum number $M$ divided by the period,
$m=M/(n+1)$. Interestingly, the phases with odd $n$ all have $M=\SHS=2$,
including the $n=7$ and $n=9$ phases not resolved in Fig.~\ref{fig.PDAFM4}.
This of course corresponds to different \emph{magnetizations} $m$. To
understand the special significance of the value $M=\SHS$, we plot in
Fig.~\ref{fig.local9} the local expectation values of the spins, $\langle
S_i^z\rangle$, for each site of a HS chain of length $n=9$, obtained with
DMRG. The plot shows that for $0<M\le\SHS$ the \emph{odd} chain can
accomodate the finite spin by forming a N\'eel-type state. For higher $M$
this is no longer possible and spins pointing in the ``wrong'' direction
are reduced (a bulk magnon is excited).
Due to the cost in exchange energy such states are always
higher in energy than competing phases. Compare also the discussion in
Sec.~\ref{sec.intro}.

The dimer ($n=2$) phase
always consists of singlets, $M=0$. This shows that it is
energetically favorable to replace the dimer state by a state with odd $n$
or the HS state, instead of having $M>0$ for the dimers.

Finally, we turn to the pure HS phase.
At $T=0$ the system is equivalent to an infinite
antiferromagnetic $S=\SHS$ chain. We thus expect the magnetization to rise
continuously with increasing magnetic field $h$ and to reach its maximum
value $m=\SHS$, i.e., full spin alignment, at a \emph{continuous} phase
transition. Since we approximate the HS phase by the $n=99$ phase, the
continuous increase is replaced by small steps.
The position of this transition is determined by equating the energies per
site for $M=n\SHS$ and $M=n\SHS-1$. From Eq.~(\ref{eq5.eps06}) we thus
obtain the critical field
$h_c = \Delta e_n^0(2n)-\Delta e_n^0(2n-1)$,
which is proportional to $J$ and independent of $B_0$ and $V$. For $n=99$ exact
diagonalization yields $h_c \approx -7.9980\,J$, compared to the
exact result for an infinite chain,
$h_c=-4J\SHS=-8\,J$. To find the critical behavior close to this transition we
define the deviation of the magnetization from its maximum by $\Delta m \equiv
\SHS - m$. Plotting $\Delta m^2$ vs.\ $h$ (not shown) we find that $\Delta m^2$
is linear in $h_c-h$ so that the critical exponent of $\Delta m$ with respect to
the field $h$ is mean-field-like, $\beta=1/2$.

The previous discussion shows that by restricting the DMRG calculations to
$n<100$ we make an error for the transition to full spin alignment of the
order of $0.03$\%. As another way to estimate the errors, we have
determined the triple point between LS, HS, and dimer phases in
zero field and compared the result to the ``exact'' triple point shown in
Fig.~\ref{fig.PDAFM1}. The error is of the order of $0.2$\% for $V$ and
$0.1$\% for $J$.

\begin{figure}[tbh]
\centerline{\includegraphics[width=3.30in,clip]{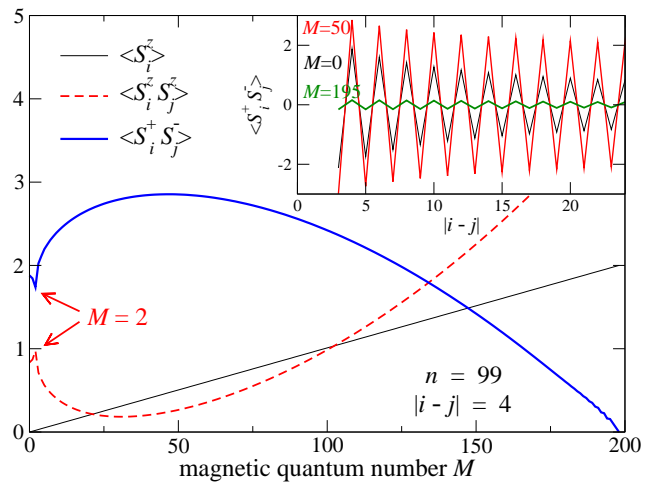}}
\caption{(Color online). Expectation value of the $z$-component of the
spins, $\langle S^z\rangle$, and spin-spin correlation functions at the
separation of $|i-j|=4$ as functions of the total magnetic quantum number
$M$ for chain length $n=99$.
Inset: Correlation function $\langle S^+_iS^-_j\rangle$ as a function of
separation for three values of the total magnetic quantum number $M$.}
\label{fig.correl}
\end{figure}

We also obtain spin correlations from the DMRG. Figure \ref{fig.correl}
shows spin-spin correlation functions for $n=99$ for two spins close to the
center of the chain, where the infinite chain should be well approximated.
We first notice the anomaly at $M=2$. This is of the same origin as the
stabilization of $M=2$ for small odd chain lengths, discussed above. It is
thus a finite-size effect not present for the true HS phase. Apart from
this anomaly, the \emph{transverse} correlations $\langle
S^+_iS^-_j\rangle$ first grow with magnetic field $h$ or magnetization.
This is the one-dimensional analog of the spin-flop state in ordered
antiferromagnets, where the staggered magnetization is oriented
perpendicularly to the applied field. For large fields, the correlations
decrease again since the spins are more and more forced into the field
direction. In the fully polarized state for $h\ge h_c$ the transverse
fluctuations vanish.

\begin{figure}[tbh]
\centerline{\includegraphics[width=3.35in]{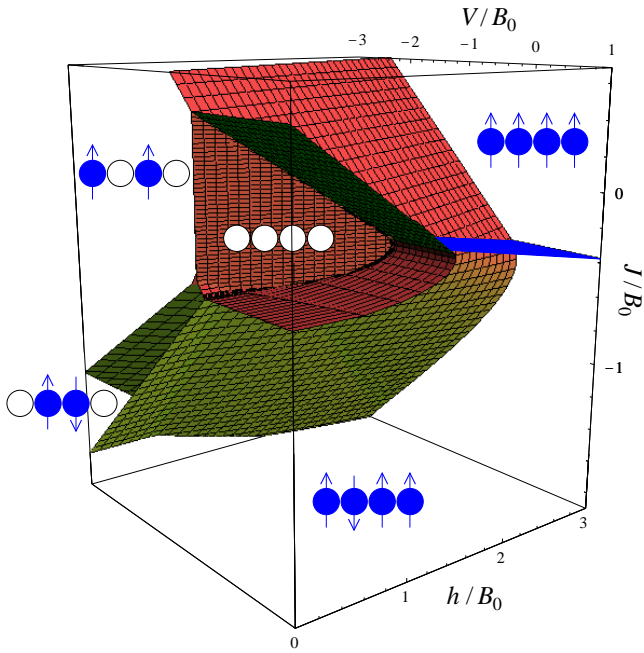}}
\caption{(Color online). Zero-temperature phase diagram of the
spin-crossover chain with $\SLS=0$ and $\SHS=2$. Positive (negative) $J$
corresponds to ferromagnetic (antiferromagnetic) exchange interaction.
$V$ denotes the elastic interaction, $h$ the applied magnetic field in units of
energy, and $2B_0$ is the energy difference between HS and LS states in the
absence of interactions. The
solid surfaces denote phase transitions between different phases, which are
indicated. The phases with HS chain lengths $n=3,5,\ldots$ are hidden in
this view. All transitions are discontinuous, except for the continuous
transition to full spin alignment in the HS phase, shown as the monochrome
(blue) surface.}
\label{fig.PD3D}
\end{figure}

\section{Summary and Conclusions}

To conclude, we have studied a model for spin-cross\-over compounds forming
one-dimensional chains. We consider spin quantum numbers appropriate for
$\mathrm{Fe}^{2+}$ ions, which have a spin-0 LS and a spin-2 HS state. The
model includes elastic and exchange interactions and an applied magnetic
field. The most important effect left out here is probably the dipole-dipole
interaction, which is of long range for an isolated chain, but becomes screened
if the chain is deposited on a conducting substrate.
We obtain the ground-state phase diagram analytically for
ferromagnetic or zero exchange interaction and using the density-matrix
renormalization group (DMRG) for antiferromagnetic exchange.
As a summary, Fig.~\ref{fig.PD3D} shows the full phase
diagram. The continuous transition to full spin alignment is indicated by
the solid blue surface. All other surfaces are discontinuous transitions
between states with different chain length $n$. Horizontal cuts correspond
to the plots in Fig.~\ref{fig.PDAFM2}, the vertical cut at $h=0$
to Fig.~\ref{fig.PDAFM1}. Besides a
diamagnetic LS phase and a HS phase equivalent to the usual Heisenberg
chain we find a number of more complex phases. For sufficiently negative
elastic interaction $V$ we find an alternating phase of low and high spins.
In quasi-two-dimensional SCC's the corresponding checkerboard state
has been observed experimentally.\cite{Co}

For antiferromagnetic coupling we find a robust dimer ($n=2$) phase, which
consists of spin singlets formed by two high spins separated by single low
spins. Since this phase appears at zero and low magnetic fields, it should be
accessible experimentally. As the magnetic field is increased, the dimers remain
in the singlet state until states with an odd number of HS ions or the pure HS
state become lower in energy, whereupon the dimer phase is destroyed in a
discontinuous transition. At higher magnetic fields we find a number of phases
consisting of finite chains of length $n=3,5,7,9$ of HS ions with total $S^z$
quantum number $M=2$ separated by single LS ions. We suggest that the succession
of odd chain lengths is a general feature of spin-crossover chains.

We thus find that a model that contains the most important ingredients of
one-dimensional spin-crossover systems shows a rich ground-state phase diagram.
The model is related to various systems studied in recent years, such as
site-diluted spin models and finite antiferromagnetic Heisenberg chains.
Questions for the future concern the behavior at nonzero temperature and of
higher-dimensional models, in which \emph{percolation} plays an important role.
New physics comes into play since the dilution by LS ions is not quenched
disorder but a dynamical degree of freedom.

\vspace*{1ex}

\acknowledgments

We would like to thank P. G\"utlich, P. J. Jensen, and F. von Oppen for valuable
discussions. C.T. thanks the Deutsche For\-schungs\-ge\-mein\-schaft for support
through Son\-der\-for\-schungs\-be\-reich 290.


\begin{thebibliography}{99}
\bibitem{rem.sigmaop}Here, $\sigma_i$ is a quantum number, not an
operator, but it is permissable to use the same symbol since the operator
$\sigma_i$ commutes with any other operator.
\bibitem{rem.degen}The entropy change associated with
the LS to HS transition is larger than predicted from these
degeneracies due to the change in the vibrational
properties between LS and HS states and can be taken into account
by effective degeneracy factors.\protect\cite{BoK95,GKA00}
\bibitem{rem.largeM}For $M=n\SHS-2$ and $M=n\SHS-1$ the results have been
obtained by exact numerical diagonalization. For $M=n\SHS$ the exact analytical
value $\Delta e_n^0(n\SHS)/|J| = (n-1)\SHS^2$ has been used.
\end{thebibliography}
\end{document}